# Determinação do centro de massa de uma peça triangular por meio de partições a partir da mediana

# Determination of the center of mass of a triangular piece by means of partitions from the median

José Joaquín Lunazzi[1]
Bruno Fontes de Sousa[2]

17 de abril de 2021

**Resumo**
É conhecido o método de se determinar o centro de massa de uma peça triangular pendurando-a por cada um de seus vértices e tracejando as verticais. O ponto de encontro das três verticais é considerado o centro de massa, o que se verifica colocando a peça em equilíbrio sobre um elemento pontiagudo. O que nem todos sabem é que esse método foi desenvolvido pelo sábio grego Arquimedes há 2.300 anos, e que ele o liga à Fig. geométrica das medianas de um triângulo. Ele demonstrou o resultado teoricamente e, inspirado na sua demonstração temos desenvolvido uma outra que conduz à ideia e métodos do cálculo diferencial.
**Palavras-chave:** Ensino de Física, Centro de massa, Equilíbrio.

**Abstract**
It is well known the method of determining the center of mass of a triangular piece in which it is hanged from each one of its vertices while drawing from the vertice its verticals. The intersection of the three verticals is considered as the center of mass, verified by equilibrating the piece over a point object. But not everybody knows that the method was developed by Arquimedes 2.300 years ago, determining the geometrical elements to the medians of the triangle. He demonstrated theoretically its result and, inspired by his demonstration we developed another one, which lends to the idea and methods of the differential and integral calculus.

## 1   Introdução

De acordo com [1], o ensino da disciplina Cálculo Diferencial e Integral I em cursos de licenciatura em Matemática apresenta "uma grande tensão entre o rigor e a intuição, exigindo do estudante uma capacidade de abstração que, muitas vezes, ele ainda não desenvolveu". Por outro lado, os autores de [2] afirmam que "o ensino da disciplina Cálculo Diferencial e Integral (CDI) desponta como um dos principais desafios". Além disso, os autores ressaltam que:

> Parte dos pesquisadores em educação matemática atribui o problema à metodologia tradicional de ensino que induz uma postura não crítica nos estudantes, pouca participação, e a não compreensão dos conceitos devido a não contextualização do ensino [2].

Em nosso ponto de vista, isto comprova a dificuldade que muitos alunos encontram para compreender os conceitos desta disciplina. Entendemos que para

---

[1]   Universidade Estadual de Campinas (UNICAMP). Instituto de Física
[2]   Universidade Federal Rural do Semi-Árido (UFERSA). Doutorando PECIM/UNICAMP

os alunos de carreiras não especificamente matemáticas seria melhor partir de conceitos mais ligados ao cotidiano dos elementos que o rodeiam, seguindo a pedagogia de Paulo Freire [3]–[9]. Desenvolvemos uma sequência paralela à da história da humanidade, onde Isaac Newton [10], [11] e Gottfried Wilhelm Leibniz [10], [12], foram os primeiros que descobriram independentemente o Cálculo [10], [13]. Arquimedes, com suas engenhosas demonstrações, resolveu problemas de Mecânica antes mesmo do surgimento do CDI, incluindo até perfis parabólicos, pelo que deve ser considerado um predecessor importante [14], [15].

Encontrar o ponto de equilíbrio de objetos com formas geométricas básicas pode não ser tão simples e levar a assumir soluções não demonstradas nem com demonstração justificada na bibliografia. Isto pode acontecer até mesmo em dissertações acadêmicas [16], [17].

De acordo com [14] é possível encontrar o ponto de equilíbrio de uma peça triangular pelo método da suspensão pelos vértices. No contexto, o autor descreve como Arquimedes provou a localização do centro de massa usando duas demonstrações por caminhos diferentes. Na primeira delas, ele traça uma mediana e linhas paralelas equidistantes a partir de pontos que resultam da subdivisão da base. Acompanhando esta demonstração pensamos em partir da figura de um triângulo escaleno para eliminar toda simetria inicial. Ao longo do processo, percebemos que a demonstração podia ser realizada aumentando o número de elementos da partição para chegar a poder desprezar os elementos menores, constituindo um caminho que facilita a ideia do uso do CDI e a própria demonstração do método clássico de determinação do centro de massa aplicado a objetos triangulares [18].

## 2    Centro de massa de uma peça triangular

Arquimedes estabelece inicialmente sete postulados. Utilizamos três deles [6], e os interpretamos assim:

**Postulado 1.** "Postulamos que pesos iguais se equilibram a distâncias iguais (Fig. 1a) e que pesos iguais suspensos a distâncias desiguais não se equilibram, mas que se inclinam do lado do peso suspenso à maior distância (Fig. 1b)."

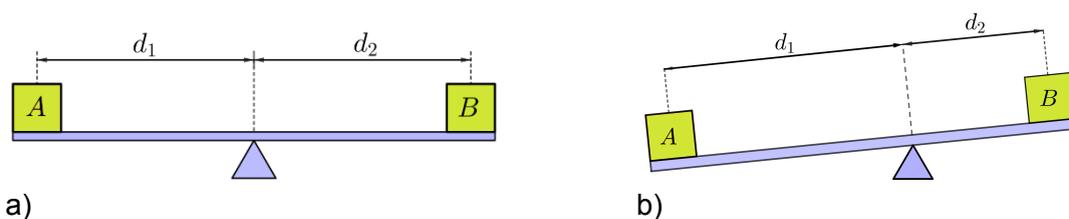

a)                                                                                     b)
**Figura 1**: a) Pesos iguais e distâncias iguais. b) Pesos iguais e distâncias diferentes

**Postulado 2.** "Quando pesos suspensos a certas distâncias estão em equilíbrio (Fig. 2a), se adicionarmos algum corpo a um dos dois pesos, os pesos não mais se equilibrarão, mas haverá uma inclinação do lado do peso ao qual foi adicionado aquele corpo (Fig. 2b)."

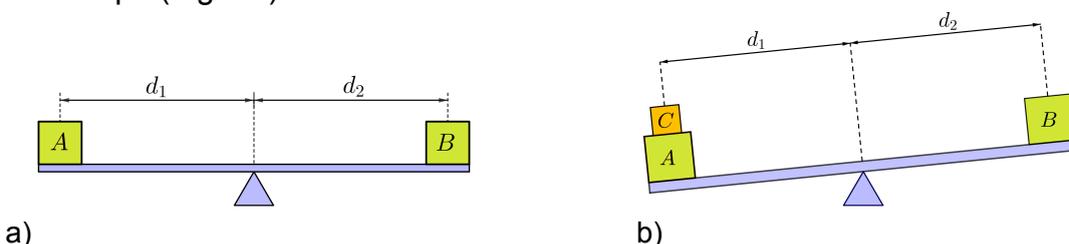

a)                                                                                     b)
**Figura 2**: a) Pesos iguais e distâncias iguais. b) Adiciona-se um peso em um dos lados da balança.

**Postulado 3.** "Da mesma forma, se removermos qualquer parte de um dos dois pesos que se equilibravam a certas distâncias (Fig. 3 a 5), os pesos não mais se equilibrarão, mas haverá uma inclinação do lado do peso do qual nada foi retirado (Fig. 3b)."

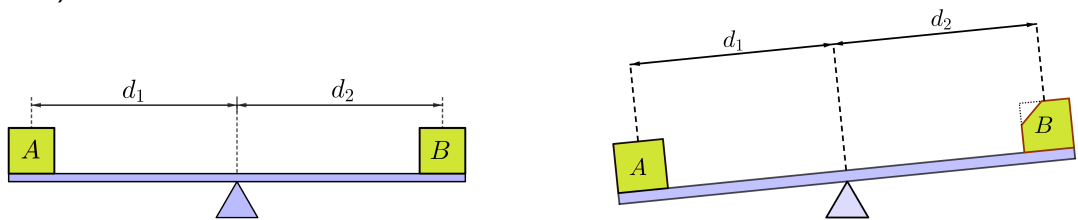

**Figura 3:** a) Pesos iguais e distâncias iguais. b) Retira-se um pouco do peso de um dos lados

Estendemos estes postulados aos casos da Fig. 4, que colocamos na consideração do leitor para ele concluir que as situações de equilíbrio permanecem sempre que houver conexões rígidas e de pesos desprezíveis.

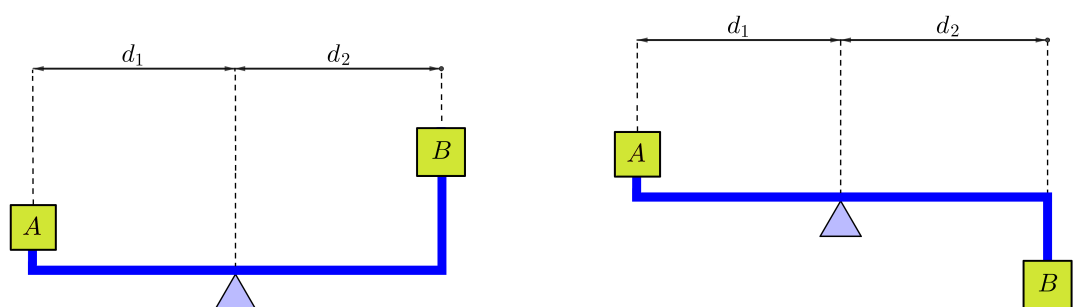

**Figura 4:** Diferentes posições verticais dos pesos.

Como consequência, observando a Fig. 5, se entende que ela é também um caso de equilíbrio. Isto nos leva a entender o conceito de braço de torque, que nas figuras são as distâncias $d_1$ e $d_2$.

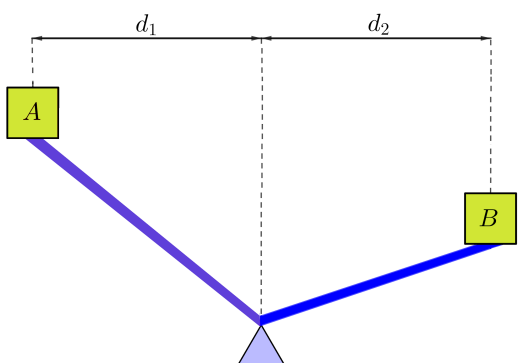

**Figura 5:** Equilíbrio com braços de torque iguais.

Na ação de uma força uniforme em todos os elementos do corpo, o raciocínio que foi colocado para o peso, vale para a massa uniforme dele, pela segunda Lei de Newton, e passaremos a usar a expressão centro de massa, como forma mais geral, ao invés de centro de peso. Passamos a analisar o caso do centro de massa de um triângulo acrescentando uma proposição de Arquimedes.

**Proposição 1.** Em todo triângulo, o centro de massa está situado sobre uma mediana [6].

**Demonstração.** Seja $ABC$ um triângulo qualquer, conforme a Fig. 6, traçamos a mediana relativa ao vértice $A$, onde $D$ é o ponto médio do lado $BC$.

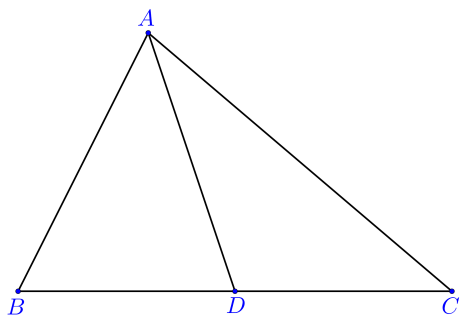

**Figura 6:** Mediana do triângulo escaleno.

Arquimedes utiliza o raciocínio de demonstração por redução a um absurdo. Assim, o centro de massa do triângulo $ABC$ estaria localizado em algum ponto do triângulo, mas fora da mediana $A$, sendo $X$ o suposto centro de massa (Figura 7).

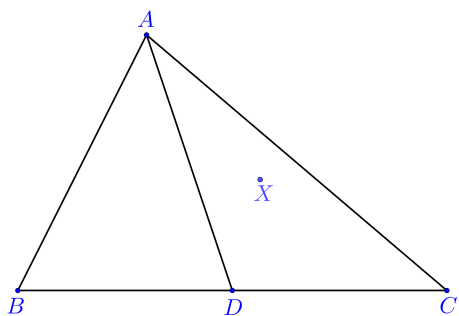

**Figura 7:** Localização hipotética do centro de massa.

Vamos demonstrar que o centro de massa do triângulo $ABC$ está situado sobre a mediana $AD$, adaptando as ideias que Arquimedes usou em sua demonstração. Começamos dividindo a base $BC$ em partes iguais na Figura 8, usando pontos equidistantes, e traçando por estes pontos segmentos paralelos à mediana $AD$.

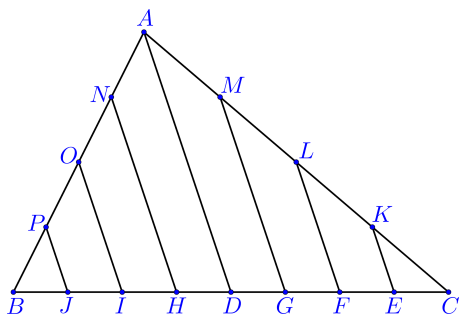

**Figura 8:** Partição inicial do triângulo.

Traçamos os segmentos $MN$, $LO$ e $KP$, paralelos à base $BC$ (Figura 9).

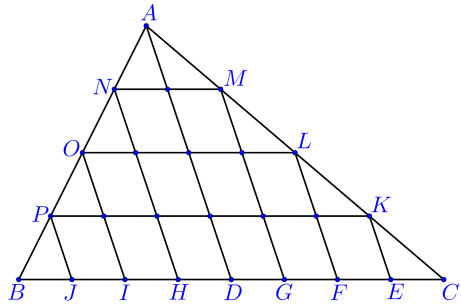
**Figura 9:** Partição horizontal do triângulo.

A Figura 10 apresenta os centros de massa dos paralelogramos determinados pelos segmentos paralelos à mediana e pelos segmentos paralelos à base do triângulo. Arquimedes [6] provou que a localização do centro de massa de cada paralelogramo correspondia ao cruzamento das linhas ligando os pontos médios dos lados opostos. No anexo 1, acrescentamos a demonstração de que a localização do centro de massa de um paralelogramo também é a interseção das diagonais. Como ilustramos no paralelogramo que contém o vértice $N$.

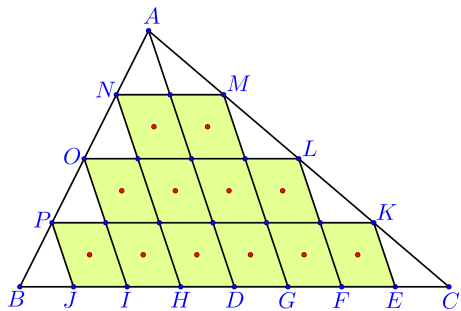
**Figura 10:** Centros de massa das partições realizadas.

A Figura 11 apresenta a peça pendurada por um de seus vértices, onde o vetor $\vec{g}^*$ representa a vertical do lugar, como se fosse um prumo. Indicamos com ele a direção do peso, dada pela gravidade, pela aceleração da rotação da Terra, pelo empuxo do ar e por mais algum outro elemento que possa intervir. **Não se pode deixar de lado o fato de que a Terra gira, reduzindo em 3% aproximadamente o peso, e fazendo com que a vertical do lugar não aponte ao centro dela**.

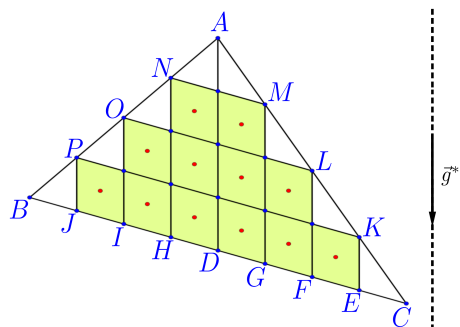
**Figura 11:** Placa pendurada por um vértice.

Note-se que a placa pendurada pelo ponto $A$ tem, do outro lado da mediana, para cada paralelogramo um que atua com torque igual e contrário. Isto porque os braços de torque de um e outro são iguais, sendo eles a distância à mediana.

Faltaria provar que o equilíbrio também corresponde se incluirmos as partes triangulares ainda não consideradas. Arquimedes o faz de uma maneira muito elaborada, e nós também poderíamos o fazer aplicando mais geometria. Vamos, porém, utilizar um outro raciocínio que vai nos permitir entender o encaminhamento ao CDI. Ele consiste em aumentar o número de partições. Dessa maneira vemos que a área total dos setores triangulares resulta cada vez menor, ficando evidente que poderemos chegar a desprezá-la. Fazendo o número de partições tender a infinito, a área desconsiderada tende a zero. Provamos assim o equilíbrio da placa em relação a qualquer ponto de sua mediana. Repetindo o processo para qualquer um dos dois vértices restantes, localizaremos um único ponto comum às três medianas [19]. Este ponto comum às três medianas constituem o que chamamos de "centro de massa".

## 3   Equilíbrio por um ponto

Note-se que, na Fig. 11, se pendurássemos a peça por qualquer ponto da mediana, o equilíbrio dos torques permanece. Se apoiarmos a placa sobre a ponta de um lápis no ponto de interseção das três medianas, e tomarmos um elemento de peso qualquer veríamos que ele tem um complemento o equilibrando a respeito de cada uma das três medianas. Isto prova que a placa resulta equilibrada? Esta é uma pergunta interessante para você, leitor, refletir. Sobre o tema, estamos elaborando um novo artigo, a ser publicado [20].

## 4   Conclusão

Mostramos de maneira didática e também histórica a justificativa para o experimento simples tão comum de determinação do centro de massa de um triângulo pendurado pelos vértices. Preenchemos desta maneira uma carência dos livros didáticos. Indicamos um raciocínio que ajuda ao aluno a aplicar a ideia de partições em grau crescente para chegar a um resultado exato, que é fundamento do cálculo diferencial e integral. Outra carência dos livros didáticos que temos identificado está em não considerar a rotação da Terra na hora de avaliar o peso. Esse erro conceitual está muito presente no cotidiano do ensino de física.

Este trabalho leva a considerar o caso mais geral de equilíbrio a respeito de qualquer reta que passe pelo centro de massa, que também pode ser desenvolvido com a mesma metodologia.

A ideia de infinitésimos que conceituamos neste artigo por meio da geometria é paralela à ideia algébrica de limite no CDI.

Ensino de Física. A rede social mundial gratuita de pesquisadores Research Gate, que possibilita a leitura de artigos publicados e de projetos e a comunicação entre os membros.

**Referências**

## Anexo 1

A seguir, vamos apresentar a definição de triângulos iguais.

**Definição 1.1** Um triângulo é igual[3] (ou congruente, símbolo ≡) a outro se, e somente se, é possível estabelecer uma correspondência entre seus vértices de modo que:

1. Seus lados são ordenadamente iguais aos lados do outro e
2. Seus ângulos são ordenadamente iguais aos ângulos do outro. [22]

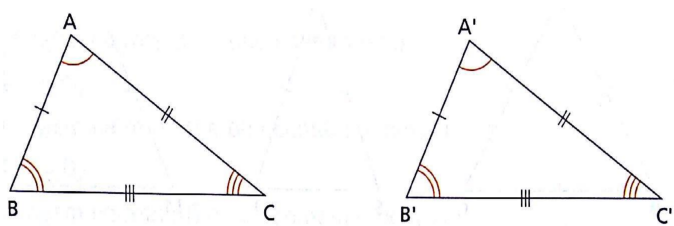

**Figura 1.1:** Igualdade de triângulos.

Os triângulos $ABC$ e $A'B'C'$ são iguais (em símbolos: $ABC \equiv A'B'C'$) se, e somente se: 1) $AB \equiv A'B'$, $AC \equiv A'C'$ e $BC \equiv B'C'$ e 2) $A \equiv A'$, $B \equiv B'$ e $C \equiv C'$.

**Proposição 1.1** Em todo paralelogramo o ponto de interseção das diagonais é o ponto médio destes segmentos.

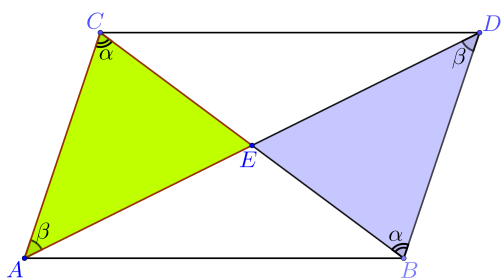

**Figura 1.2:** Igualdade dos triângulos $ACE$ e $DBE$.

**Demonstração.** Demonstramos que os triângulos $ACE$ e $DBE$ da Fig. 1.2 são iguais, pelas seguintes relações:

1. $E\hat{C}A \equiv E\hat{B}D$ e $E\hat{A}C \equiv E\hat{D}B$, por serem alternos internos de retas paralelas.
2. $BD \equiv BD$, por serem lados opostos do paralelogramo.

---

[3] Usaremos a palavra "igual" ao invés do termo matemático "congruente", para facilitar a compreensão. Assim, diremos que dois triângulos são iguais se satisfazem a definição 1.1.

Daí, concluímos que $AE \equiv ED$ e $CE \equiv EB$.

**Definição 1.2** (Ponto como elemento físico). Consideramos cada ponto como uma partícula de massa e dimensões mínimas, iguais para todos.

**Proposição 1.2.** Em todo paralelogramo, o centro de massa (de massa, ou de equilíbrio) é o ponto de interseção das diagonais.

**Demonstração.** Pela lei da alavanca, aplicada nas Fig. 1.3, os pontos indicados por $P$ e $Q$ estarão em equilíbrio se o paralelogramo estiver na horizontal e apoiado unicamente no ponto $E$ (ponto de interseção das diagonais do paralelogramo $ABCD$).

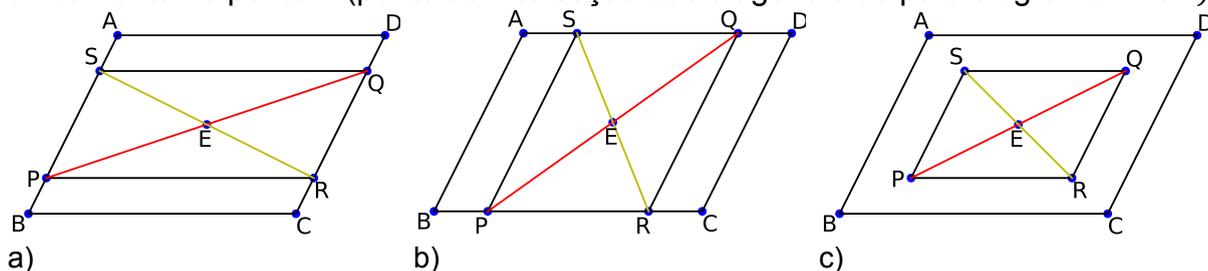

Figura 1.3: a) $P$ em $AB$. b) $P$ em $BC$. c) $P$ no interior do paralelogramo